*Article*

# Configurational Information as Potentially Negative Entropy: The Triple Helix Model

**Loet Leydesdorff**

Amsterdam School of Communications Research (ASCoR), University of Amsterdam, Kloveniersburgwal 48, 1012 CX Amsterdam, The Netherlands. E-Mail: loet@leydesdorff.net; Website: http://www.leydesdorff.net

**Abstract:** Configurational information is generated when three or more sources of variance interact. The variations not only disturb each other relationally, but by selecting upon each other, they are also positioned in a configuration. A configuration can be stabilized and/or globalized. Different stabilizations can be considered as second-order variation, and globalization as a second-order selection. The positive manifestations and the negative selections operate upon one another by adding and reducing uncertainty, respectively. Reduction of uncertainty in a configuration can be measured in bits of information. The variables can also be considered as dimensions of the probabilistic entropy in the system(s) under study. The configurational information then provides us with a measure of synergy within a complex system. For example, the knowledge base of an economy can be considered as such a synergy in the otherwise virtual (that is, fourth) dimension of a regime.

## 1. Introduction

When variation is considered as a relative frequency or probability distribution ($\Sigma_i\, p_i$), the Shannon-type information or the uncertainty contained in the distribution ($H$) is defined [1, 2] as follows:

$$H_i = -\Sigma_i\, p_i \log_2 (p_i) \tag{1}$$

Equivalently, for a two-dimensional distribution (e.g., a matrix), $H_{ij}$ is:

$$H_{ij} = - \Sigma_i \Sigma_j \, p_{ij} \log_2 (p_{ij}) \qquad (2)$$

This uncertainty is the sum of the uncertainty in the two dimensions of the probabilistic entropy diminished by their mutual information. In other words, the two variations overlap in their co-variation, and condition each other asymmetrically in the remaining variations (Figure 1).

**Figure 1.** Relations of expected information contents, mutual information, and conditional entropies between two variables *x* and *y* [3].

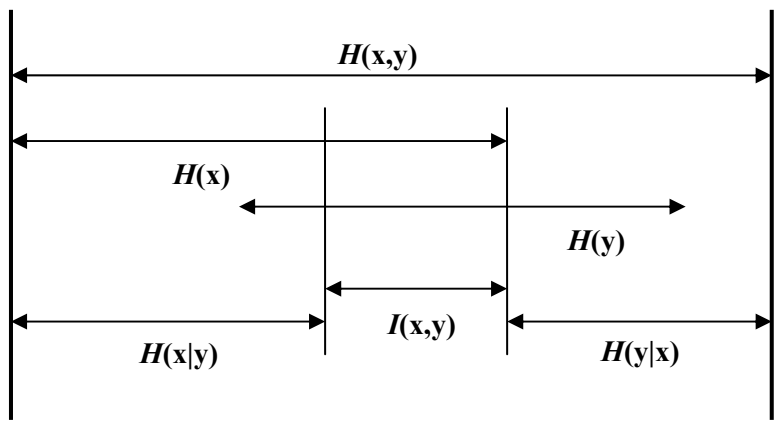

Mutual information has been formalized in information theory using the concept of *transmission* [3-5] in the case of two variables, but more generally using the *I*-measure $\mu^*$ [6: 53 ff.]:

$$H_{ij} = H_i + H_j - I_{ij} \qquad (3)$$
$$I_{ij} = H_i + H_j - H_{ij} \qquad (4)$$

$I_{ij}$ is zero if the two distributions are completely independent (i.e., the co-variation is zero), but otherwise necessarily positive [5]. The *I*-measure $\mu^*$ is signed because as we shall see below the uncertainty may be reduced in the case of more than two variables.

**Figure 2.** The mutual information between two systems may endogenously develop a third dimension of the probabilistic entropy [7].

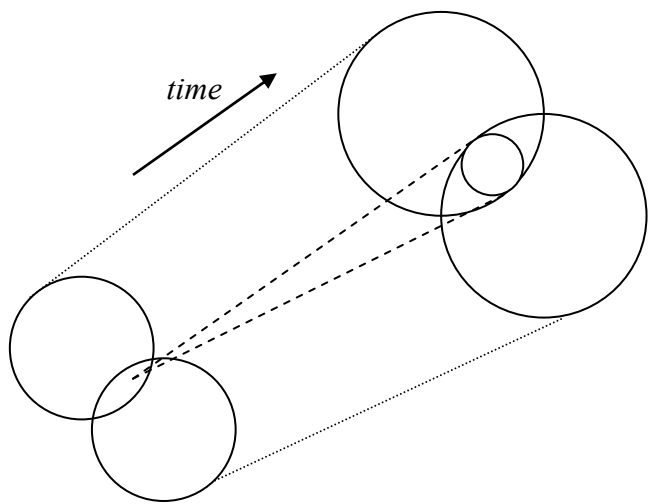

Figure 2 depicts how a two-dimensional system can develop into a three-dimensional one. Over time, the mutual information may be reproduced increasingly as a potentially separate system. A system, for example, can be defined by using the Markov property: if the current state of a distribution provides a prediction for the system's next state better than the sum total of the development of its elements (after proper normalization), an emerging axis of systemness can be expected [7-9].

The configurational information in three dimensions of the probabilistic entropy can be defined as follows [4: 129, 6: 55 ff.]:

$$\mu^*_{ijk} = H_i + H_j + H_k - H_{ij} - H_{ik} - H_{jk} + H_{ijk} \tag{5}$$

The resulting indicator can be negative or positive (or zero) depending on the relative sizes of the contributing terms. A negative value means that the uncertainty prevailing at the network level is reduced. McGill [10, 11] proposed calling this reduction of uncertainty "configurational information" [12]. Because this information is configurational, the reduction of uncertainty cannot be attributed to one single contributor. These network effects are systemic [13]. Note that the bilateral terms contribute to the reduction of uncertainty, while uncertainty in three dimensions adds to the uncertainty which prevails.

The extension of the *I*-measure in three dimensions to more dimensions is straightforward. In the case of four dimensions, for example, one can formalize as follows:

$$\mu^*_{ijkl} = H_i + H_j + H_k + H_l - H_{ij} - H_{ik} - H_{il} - H_{jk} - H_{jl} - H_{kl} \\ + H_{ijk} + H_{ijk} + H_{jjl} + H_{ikl} - H_{ijkl} \tag{6}$$

With each extension of another dimension another sign is added. In the first-order variation, redundancy as the complement of uncertainty is specifiable in bits of information. In a two-dimensional probability distribution or matrix, a latent eigenstructure can be hypothesized. I shall argue that in the case of three-dimensional probability distributions, arrangements can compete along trajectories for globalization. A globalizing regime can be expected to remain pending as selection pressure on relatively stabilized configurations. I suggest reserving the word "latent" for structure (at each moment of time) [14], and to use the word "virtual" for a regime developing over time [15-17].

## 2. Emerging Configurations

Let me emphasize that despite the natural-science connotation of the word entropy, entropy statistics remains a purely mathematical tool. If two is the basis of the logarithm in Equations 1-6, probabilistic entropy is measured in (dimensionless!) bits of information. The measurement results have yet to be provided with meaning in terms of the system(s) under study. Because all terms in information theory are composed of summations (using Σs), the statistical decomposition can be pursued to the level of specifying the contribution of individual cases to the uncertainty [5, 18]. In other words, this methodology enables us to evaluate higher-dimensional configurations in detail, and also to specify the effect of adding another dimension to our model.

Shannon detached himself from the implications of his counter-intuitive definition of information as uncertainty by declaring that the 'semantic aspects of communication are irrelevant to the engineering problem' [1, 2]. The mathematical theory of communication has often been discredited in the social sciences with the argument that Shannon defined information as uncertainty [19]. This definition is counter-intuitive, since one is inclined to associate 'information' with a message that 'in-forms' a receiving system [20]. Shannon-type information, however, is defined as content-free (that is, in terms of bits of information) and without meaningful dimensionality. Therefore, this mathematical apparatus can be applied to any (change in) probability distribution(s).

The dimensionless-ness and therefore meaninglessness of Shannon-type information finds technically its origin in using only the probabilistic part of the so-called Gibbs entropy formula, which is formulated as follows:

$$S = -k_B \sum_i p_i \log p_i$$
$$= k_B H$$

The Boltzmann constant ($k_B$) provides dimensionality (Joule/Kelvin) to the entropy $S$, whereas the Shannon-entropy $H$ is based on only the probabilistic part of the equation. In the case of a thermodynamic system, this is typically the 6$N$-dimensional space of $N$ particles. In other words, the specification of a system of reference—in this case, a thermodynamic system—provides the Shannon-type information with meaning. In general, this specification of a system of reference (potentially different from a thermodynamic system) also makes it possible to specify the maximum entropy $H_{max} = \log(n)$ in which $n$ is the *number* of categories.

In other words, Shannon-type information itself may not *yet* imply meaning, but the emphasis is on the "yet." The specification of maximum entropy requires the specification of a system, and thereby (that is, by the substantive application of the mathematical apparatus) a specific meaning is provided to the Shannon-type information by the researcher. In other words, the specific meaning of Shannon-type information is determined by the substantive theories used in the research design [7]. For example, the probabilistic entropy contained in a transaction matrix informs us about this matrix in terms of bits of information [5]. The expected information content of the matrix is the complement of the redundancy in this probability distribution since the maximum information is by definition the sum of these two terms.

From the perspective of systems theory, the maximum entropy of a system of $N$ agents would be $\log(N)$, but the number of possible transactions among these $N$ agents is $N*(N-1)$. In the former case, the Shannon information is expected to be contained in a one-dimensional probability distribution or vector, and only variation along this vector can be measured. Using a two-dimensional probability distribution, i.e., a matrix, one can measure both variation and structure. A matrix, for example, can be considered as a representation of a network with a specific structure. Eigenvectors of the matrix provide us with an expectation of its structure. Structure, however, operates deterministically as selection upon variation.

When the Shannon-type information is expected to be contained in a three-dimensional probability distribution, a trajectory can be shaped in the corresponding cube of information (Figure 3a). A specific network structure may be stabilized to variable extents, which then are measurable in terms of bits of information across different probability distributions. If a fourth dimension is added, a self-organizing regime can be shaped at the systems level because the system now has one more degree of freedom available for adjusting its representations of the past with the hindsight of the present (Figure 3b).

**Figure 3a.** An observable trajectory of a (potentially complex) system in three dimensions.

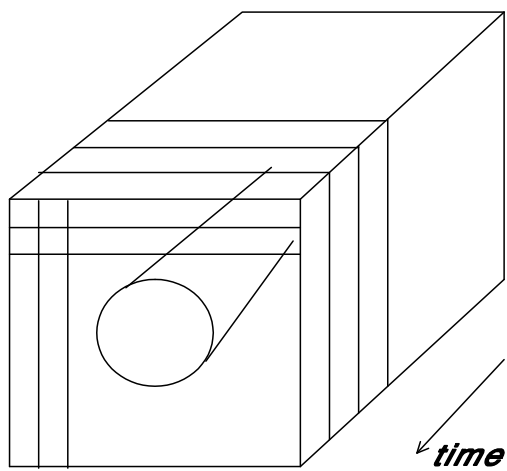

**Figure 3b.** Selection among representations of the past using a fourth degree of freedom.

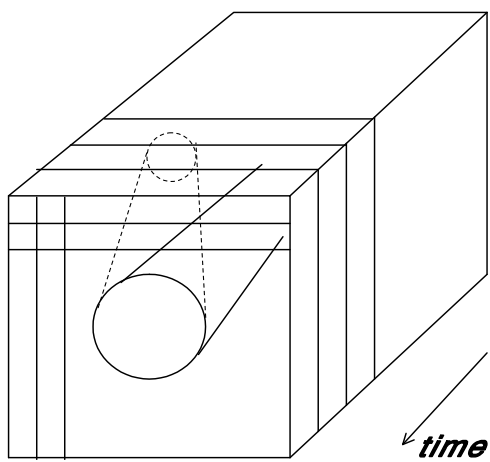

Using the perspective of evolution theory, one can also formulate that the variation is stochastic and can thus be considered as a one-dimensional vector. Selection operates in a second (orthogonal) dimension and is structural. Some selections can be selected for stabilization over time and thus a retention mechanism can be shaped. Stabilizations can be selected for globalization (by inverting the arrow of time [21, 22]). Note that the sign of the selective operation is negative: globalization can also be considered as second-order selection on stabilizations as second-order variation [17, 23]. The system is constructed bottom-up, but a next-order system may rest as a control mechanism on the lower-order ones by selecting upon the uncertainty that prevails.

The fourth dimension of the probability distribution can thus be considered as a feedback mechanism on a three-dimensional probability distribution. While the forward moving probability distribution can be expected to follow the second law [5], this feedback in a fourth dimension of the probability distribution may generate negative entropy. This negative entropy is measurable as "configurational information."

## 3. Storper's (1997) model of the knowledge base of regional economies

In *The Regional World*, Michael Storper [24] argued that technology, organization, and territory can be considered as a 'holy trinity' for regional development. Storper emphasized that this holy trinity should not be studied as an aggregate of the composing elements, but in terms of the relations between and among these elements. These relationships shape regional economics. However, Storper's proposal for a 'relational paradigm' was not yet operationalized in terms which allowed for measurement.

Using the configurational information among distributions of firms in the three relevant dimensions (territory, technology, and organization), one can measure the potential synergy at national and regional levels. In the geographical dimension, the postal codes provide a fine-grained map of regions and countries which can be classified at different levels of aggregation. We [25, 26] used the *Nomenclature of Territorial Units for Statistics* (NUTS) which was established by Eurostat more than 25 years ago in order to provide a single uniform breakdown of territorial units for the production of regional statistics for the European Union. NUTS (*Nomenclature des Unités Territoriales Statistiques*) classifications can be retrieved at http://ec.europa.eu/eurostat/ramon/nuts/home_regions_en.html. The so-called NACE-classification (*Nomenclature générale des Activités économiques dans les Communautés Européennes*) of the Organization of Economic Cooperation and Development (OECD) provided us with the technological classes of firms, and additionally allows us to distinguish classes as knowledge-intensive services and medium or high-tech industries (Table 1). Firm size in terms of numbers of employees can be used as a proxy for the industrial organization [27-29].

**Table 1**. Classification of high-tech and knowledge-intensive sectors according to Eurostat [30].

| *High-tech Manufacturing* | *Knowledge-intensive Sectors (KIS)* |
|---|---|
| **30** Manufacturing of office machinery and computers | **61** Water transport |
| **32** Manufacturing of radio, television and communication equipment and apparatus | **62** Air transport |
| | **64** Post and telecommunications |
| **33** Manufacturing of medical precision and optical instruments, watches and clocks | **65** Financial intermediation, except insurance and pension funding |
| | **66** Insurance and pension funding, except compulsory social security |
| *Medium-high-tech Manufacturing* | **67** Activities auxiliary to financial intermediation |
| | **70** Real estate activities |
| **24** Manufacture of chemicals and chemical products | **71** Renting of machinery and equipment without operator and of personal and household goods |
| **29** Manufacture of machinery and equipment n.e.c. | **72** Computer and related activities |
| | **73** Research and development |
| **31** Manufacture of electrical machinery and apparatus n.e.c. | **74** Other business activities |
| | **80** Education |
| **34** Manufacture of motor vehicles, trailers and semi-trailers | **85** Health and social work |
| | **92** Recreational, cultural and sporting activities |
| **35** Manufacturing of other transport equipment | Of these sectors, **64, 72** and **73** are considered *high-tech services*. |

We obtained data for Germany and the Netherlands, but at different levels of aggregation. In the Dutch case, the data consists of 1,131,668 records containing micro-information based on the registration of all enterprises by the Chambers of Commerce. Because registration with the Chamber of Commerce is obligatory for corporations, the dataset covers the entire population at the micro level ($N = 1,131,668$). This data is collected by Marktselect plc on a quarterly basis. Our data specifically correspond to the CD-Rom for the second quarter of 2001 [31]. We brought the data under the control of a relational database manager in order to enable us to focus on the relations more than on the attributes. Dedicated programs were developed for the further processing and computation where necessary.

The relevant statistics for Germany were generated by the Federal Employment Office (*Bundesagentur für Arbeit*) [32]. In the case of composite (e.g., international) corporations with multiple locations, the data are collected at the level of the local establishments, and thus the geographical dimension is reflected in these data. While the Dutch data was firm-data at the micro level, the German statistics (for 2002) were made available to us at the NUTS-3 level of the Eurostat classification of regions. In the German Federal Republic the NUTS-3 level coincides with the district or *Kreis* ($N = 438$). However, the data contains cross-tables with the NACE categories and numbers of employees at this fine-grained geographical level. Thus, we know how many establishments of a certain size are established in each locality for each NACE-category at the three-digit level. In order to keep the study strictly comparable with the one for the Netherlands, we used only two digits of the NACE classifications. Furthermore, the Dutch data included a category for firms with zero employment, but this category is not contained in the German statistics because self-employed persons are not obliged to report to the social insurance scheme.

Using Equations 5, the three distributions can be used for the specification of the configurational information. Additionally, Theil's decomposition algorithm [4] enables us to compare the configurational information at different levels of aggregation. Thus, the Netherlands and Germany can be considered as composed systems in terms of their lower-level units like the NUTS-2 provinces ("Regierungsbezirke") and the NUTS-3 regions ("Kreise"). (In the German case, the federal states ("Länder") are scaled as NUTS-1.) Using the decomposition algorithm, the distributions are weighted in the various dimensions for the number of firms in the subgroups *i* by summing first the uncertainties within the different groups ($\sum_i (n_i/N) * H_i$; $N = \sum_i n_i$). The in-between group uncertainty $H_0$ is then defined as the difference between this sum and the uncertainty prevailing at the level of the composed system:

$$H = H_0 + \sum_i (n_i/N) H_i \qquad (7)$$

Or equivalently for the mutual information (in the case of two variables) or the *I*-measure $\mu^*$ more generally. (The formula is equally valid for the transmission or *I*-measure because these are based on the probability distributions in the mutual information between two or more probability distributions. The probability distribution in the mutual information $I_{ab}$ can be written as the intersect between the distributions for *a* and *b,* or in formula format as $\sum p_I = \sum (p_a \text{ AND } p_b)$.)

$$\mu^* = \mu^*_0 + \sum_i (n_i/N) \mu^*_i \qquad (8)$$

Note that in this case, the regions and provinces are not compared in terms of their knowledge intensity among themselves, but in terms of their weighted contributions ($\Delta\mu^*$) to the knowledge base of the national economy as a whole. Analogously, one can decompose along the dimensions high-,

medium-, and low-tech manufacturing versus services which are knowledge-intensive and/or high-tech.

The resulting reduction of the uncertainty or configurational information can be considered as a measure of the "synergy" among the three dimensions or, in conceptual terms, of the extent to which a regional or national economy has become knowledge-based. The in-between group uncertainty [5] indicates whether the national level adds surplus levels to the sum of the regions (or not). In the case of the Netherlands [25], indeed, uncertainty was reduced at the national level (Table 2), but in the case of Germany [26] this synergy was found at the level of the federal states ("Länder").

**Table 2**. The configurational information in three dimensions for the Netherlands statistically decomposed at the NUTS 2-level (12 provinces) in millibits of information [25].

|  | $\Delta\mu^*_{GTO}$ (= $n_i * \mu^*_i$ /N) in millibits of information | $n_i$ of firms |
|---|---|---|
| Drenthe | -1.29 | 26,210 |
| Flevoland | -0.55 | 20,955 |
| Friesland | -1.79 | 36,409 |
| **Gelderland** | **-4.96** | 131,050 |
| Groningen | -1.20 | 30,324 |
| Limburg | -1.96 | 67,636 |
| **N-Brabant** | **-5.56** | 175,916 |
| **N-Holland** | **-3.28** | 223,690 |
| Overijssel | -1.98 | 64,482 |
| Utrecht | -1.86 | 89,009 |
| **S-Holland** | **-5.84** | 241,648 |
| Zeeland | -0.83 | 24,339 |
| Sum ($\sum_i P_i I_i$) | -31.10 | 1,131.668 |
| $\mu^*_0$ | **-2.46** |  |
| ***The Netherlands*** | **-33.55** | *N = 1,131,668* |

Figures 4a and 4b show the results for the Netherlands [25] and Germany [26], respectively. In the Netherlands, the NUTS-3 regions Amsterdam-Utrecht, Rotterdam-The Hague, and the environment of Eindhoven (including the northern part of Limburg) are the most knowledge-based regions. In Germany, the metropolitan regions around Munich, Hamburg, and Frankfurt are strongly indicated. Note that the former East/West divide no longer prevails at this level (NUTS-2), while it remained the main dividing line at the NUTS-1 level of German federal states (not shown here). In summary, configurational entropy can be used as a measure of the reduction of uncertainty generated by synergy among the geographical, technological, and organizational dimensions of the industries using appropriate proxies.

**Figure 4a.** Contribution to the knowledge base of the Dutch economy at the regional (NUTS-3) level. Source: [25].

**Figure 4b.** Contributions to the knowledge base of the German economy at the regional (NUTS-2) level. Source: [26].

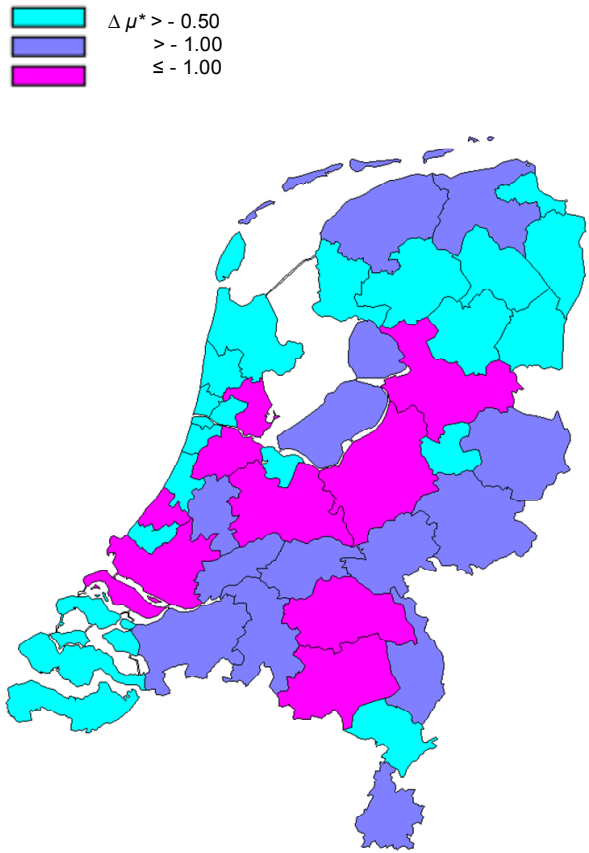
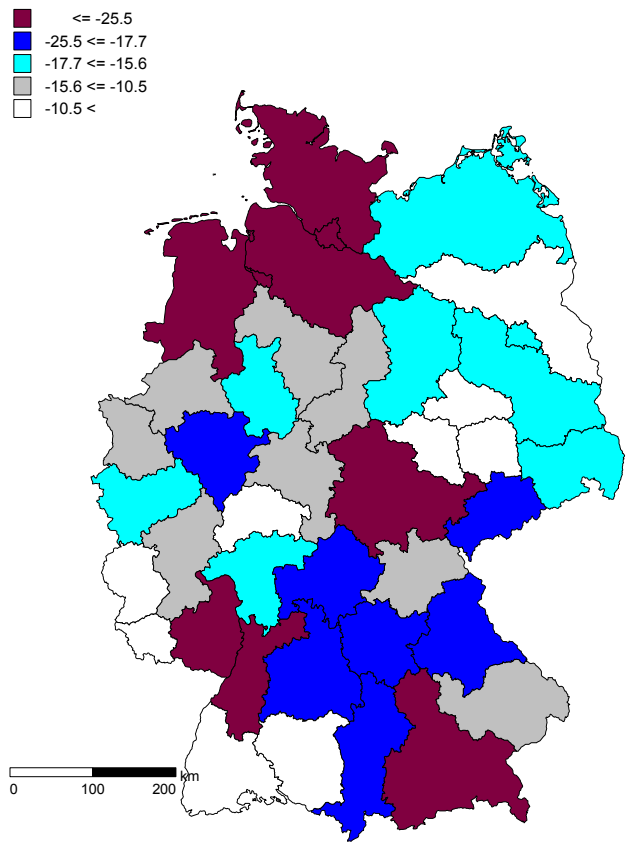

Further decomposition of the results along the sectorial axis—Table 3 for Germany, but similar results were obtained for the Netherlands—reveals that (*i*) medium-tech manufacturing provides a more important contribution to the synergy than high-tech firms, and (*ii*) knowledge-intensive services tend to uncouple the knowledge base from geographical locations. Knowledge-intensive services are not region-specific because one can take a train or plane when the service is needed elsewhere; high-tech services, however, sometimes imply the construction of a research laboratory. One can expect that high-tech firms are more globalized and therefore less embedded locally than medium-tech ones [33].

Table 4 expresses these conclusions for Germany in quantitative terms by using the respective variances as a measure of the synergetic effects. The variances clearly show that the levels of regional difference are considerably smaller for high-tech manufacturing than for medium-tech manufacturing, but as noted both have a structuring effect on the economy. The knowledge-intensive services do not contribute to structuring the knowledge-based economy differentially among regions (compared with "all sectors"), but this is less the case at the high-tech end of these services. This latter effect is enhanced in less-developed regions like Eastern Germany.

**Table 3**. *μ\** in mbits and change of *μ\**-values at the NUTS-1 level of German federal states ("Länder") [26].

| | All sectors | Manufacturing | | | | | | Services | | | |
|---|---|---|---|---|---|---|---|---|---|---|---|
| | | High-tech | Change (%)[a] | Medium-tech | Change (%)[a] | High- and medium-tech | Change (%)[a] | Knowledge intensive | Change (%)[a] | High-tech | Change (%)[a] |
| Germany | -180.08 | -163.337 | -9.3 | -202.135 | 12.2 | -199.013 | 10.5 | -161.912 | -10.1 | -164.037 | -18.8 |
| Baden-Württemberg | -47.71 | -52.672 | 10.4 | -62.893 | 31.8 | -63.281 | 32.6 | -41.168 | -13.7 | -46.155 | -3.3 |
| Bavaria | -90.41 | -110.363 | 22.1 | -142.965 | 58.1 | -137.331 | 51.9 | -67.983 | -24.8 | -79.023 | -12.9 |
| Brandenburg | -25.33 | -20.696 | -18.3 | -30.564 | 20.7 | -29.213 | 15.3 | -21.06 | -16.8 | -22.447 | -11.4 |
| Hessen | -46.24 | -44.662 | -3.4 | -58.114 | 25.7 | -56.669 | 22.6 | -36.101 | -21.9 | -35.146 | -24.0 |
| Mecklenburg-Western Pomerania | -17.68 | -10.692 | -39.5 | -22.514 | 27.3 | -22.71 | 28.4 | -16.944 | -4.2 | -19.429 | 9.9 |
| Lower Saxony | -69.3 | -72.053 | 4.0 | -94.14 | 35.8 | -94.219 | 36 | -54.223 | -21.8 | -66.376 | -4.2 |
| North Rhine-Westphalia | -49.82 | -46.531 | -6.6 | -63.765 | 28.0 | -61.668 | 23.8 | -38.333 | -23.1 | -43.291 | -13.1 |
| Rhineland-Palatinate | -53.24 | -48.53 | -8.8 | -75.113 | 41.1 | -72.715 | 36.6 | -44.538 | -16.3 | -50.665 | -4.8 |
| Saarland | -8.76 | -9.123 | 4.1 | -8.792 | 0.4 | -9.721 | 10.9 | -7.951 | -9.3 | -8.712 | -0.5 |
| Saxony | -43.03 | -40.379 | -6.2 | -58.074 | 35.0 | -56.217 | 30.7 | -31.644 | -26.5 | -36.931 | -14.2 |
| Saxony-Anhalt | -32.88 | -22.845 | -30.5 | -46.793 | 42.3 | -43.659 | 32.8 | -26.831 | -18.4 | -33.386 | 1.5 |
| Schleswig-Holstein | -40.28 | -41.331 | 2.6 | -53.345 | 32.4 | -54.838 | 36.1 | -33.144 | -17.7 | -34.15 | -15.2 |
| Thuringia | -31.12 | -25.059 | -19.5 | -45.461 | 46.1 | -43.412 | 39.5 | -22.593 | -27.4 | -29.887 | -4.0 |

[a] Percentage change is expressed with reference to "All sectors".

**Table 4**: Variances in the Δ *μ\** across knowledge-based sectors of the German economy at the NUTS-1 and NUTS-2 levels [26].

| | NUTS-1 (*Länder*) | NUTS-2 (*Regierungsbezirke*) |
|---|---|---|
| All sectors | 458.277 | 83.763 |
| knowledge intensive services | 256.176 | 57.624 |
| High-tech services | 357.369 | 75.007 |
| medium-tech manufacturing | 1133.959 | 161.110 |
| High-tech manufacturing | 744.390 | 91.898 |
| High & medium-tech manufacturing | 1058.008 | 106.112 |
| Number of regions | 13 | 38 |

## 4. The Triple Helix Model

In 1953, Linus Pauling and Robert B. Corey proposed that DNA was made up of three chains, twisted around each other in ropelike helices [34]. A few months later, James Watson and Francis Crick proposed the double helix, which was then quickly accepted as the correct structure of DNA [35]. This discovery led to a Nobel Prize [36]. Double helices can under certain circumstances stabilize in a co-evolution, but triple helices may contain all kinds of chaotic behavior [37, 38]. However, Triple Helix models continue to be useful in studying transition processes, for example, in crystallography and molecular biology.

More recently, Richard Lewontin [39] used the metaphor of a Triple Helix for modeling the relations among genes, organisms, and environments. In a different context, Henry Etzkowitz and I introduced the Triple Helix model in 1994 for studying university-industry-government relations [40-42]. University-industry-government relations provide a networked infrastructure for knowledge-based innovation systems. The infrastructure organizes the dynamic fluxes of communication through the networks. However, the resulting knowledge base remains emergent given these conditions. Whereas the relations between institutions can be measured as variables, the interacting fluxes generate a probabilistic entropy. The configurational information among the three institutional dimensions provides us with an indicator of the uncertainty at the systems level.

When the configurational information is negative, the development of a Triple Helix can be considered as self-organizing because the operation of the links in that case reduces the uncertainty in the configuration. This negative entropy cannot be attributed to any of the nodes, since it is a consequence of the configuration. However, the self-organizing dynamic of interacting expectations can be volatile or it may be stabilized in an overlay of communications which functions as a hypercycle among the carrying agencies. The dynamics of Triple Helix relations at the global and national levels in various databases and in different regions of the world can be analyzed by applying and developing the entropy indicator using scientometric and webometric data.

**Figure 5.** A schematic depiction of a complex system by Goguen and Varela [43].

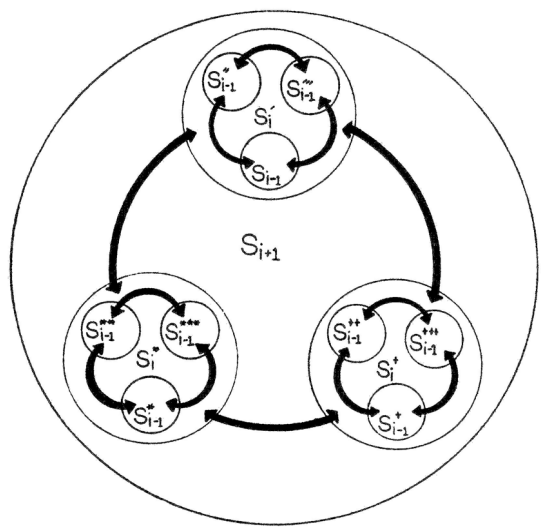

The metaphor of a Triple Helix captures the complexity that is generated when three subdynamics interact [37]. Already in 1979, Goguen amd Varela [43] proposed a representation of a complex and self-organizing system using a holographic model of three interacting subdynamics (Figure 5). At each step (*i*-1, *i*, *i*+1), the emerging system is composed of interaction effects among the previous stages of the three participating (sub)systems. In addition to the recursion of the *interaction* among the helices, however, a model of university-industry-government relations should encompass the dynamics *within* each of the helices along their respective time axes. The differences in these subdynamics (of *aggregation* and *interaction*, respectively) may break the symmetries which are suggested by the visualization in Figure 5.

How can a Triple Helix model be both interactive and recursive at the same time? First, consider three helices as sets that overlap in the intersections, as depicted in terms of Venn diagrams in Figure 6. In this configuration, the three helices share a common ground or origin in the overlap area indicated in Figure 6a as *i*. Under certain conditions, however, this overlap can also become negative. This possible configuration is depicted in Figure 6b: the center not only fails, but has become a hole or one might say a negative overlap.

**Figure 6a**. Static representation of a Triple Helix configuration as a Venn diagram.

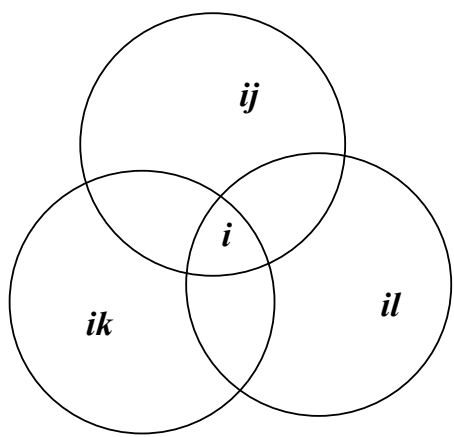

**Figure 6b**. A Triple Helix configuration with negative overlap among the subsystems.

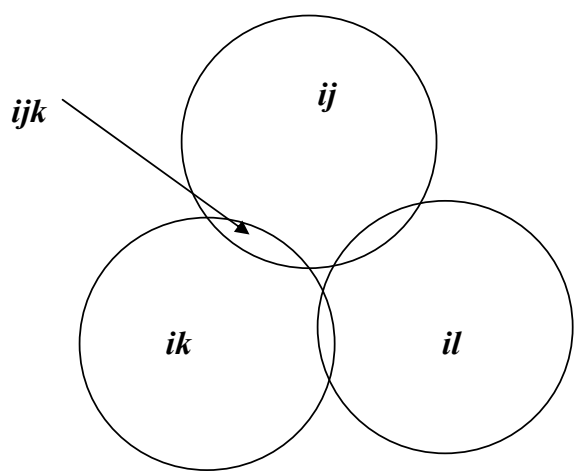

In the latter representation, the three helices have differentiated to such an extent that the commonality *i* has been dissolved. This system operates over time in terms of different communications at the respective interfaces (*ijk* in Figure 6b). If all the interfaces operate, one can consider the result as the emergence of a "hypercycle" (Figure 7). The hypercyclic configuration integrates the three systems in a distributed mode. It fails to integrate completely, or one could also say that the integration remains subsymbolic.

**Figure 7.** *Ex post* integration in an emerging hypercycle by recombining different interactions.

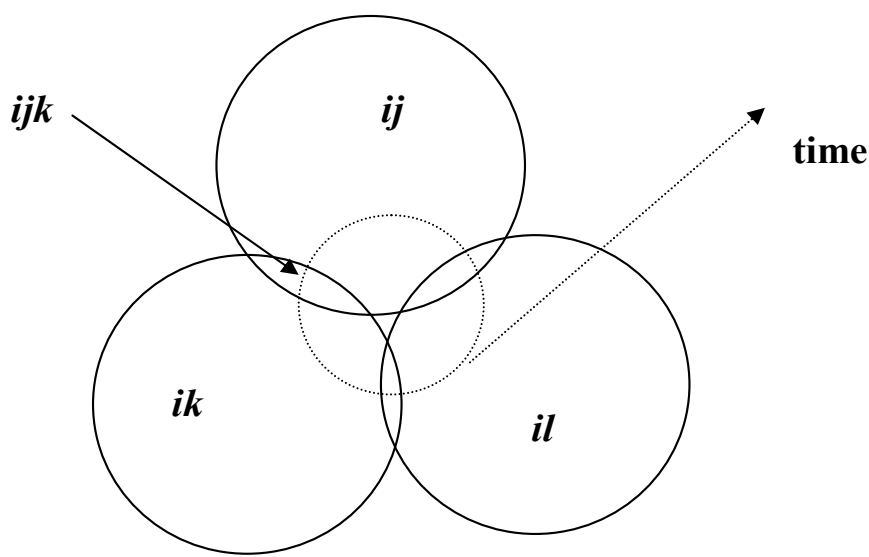

This configuration can be expected to exhibit self-organizing properties because the various transmissions are no longer integrated at a single center. A common domain of instantaneous integration is lacking, and therefore integration leads in turn to a re-differentiation. Integration fails in this configuration at each moment in time, but it can take place to variable extents over the time dimension.

Unlike integration in the previous model, this hypercyclic mode of integration is no longer controlled hierarchically by a position at the center. First, the center is now void, or one might even say negative: only an expectation of a center has remained. But more importantly, the circles in the plane and the hypercycle on top close in this configuration around a failing center. One can indicate this cavity as negative entropy within the system.

The four rings in Figure 7 can be considered as four corners of a tetrahedron as depicted in Figure 8. However, a tetrahedron can be tumbled, and thus each of the corners can become the top from the perspective of the other three cycles. The hypercycle and the various cycles stand in heterarchical relation to one another.

**Figure 8.** The hypercyclic closure of communication.

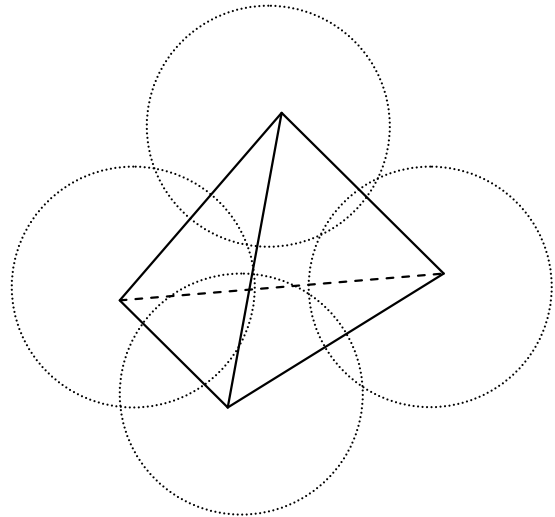

In other words, the historical structuring in terms of relations and the functional organization in terms of positions—relative to the axes of the system—operate as different subdynamics in this complex system. A system can be considered as complex when it can be decomposed into more than two subdynamics. Two subdynamics can still be modeled by extending the Darwinian model into a co-evolution model [44]. Three subdynamics may share a common zone and be integrated in a hierarchical mode given a center of control, but they can also develop into this heterarchical configuration. The hypercyclic model assumes one degree of freedom more than the configuration on which it rests. This additional degree of freedom can be used for knowledge-based anticipation in the network of relations [45].

## 5. Empirical studies of Triple Helix configurations

In several research projects we have used configurational information as an indicator in measurements of university-industry-government relations. Table 5, for example, provides values for the transmission in these relations using co-authorship relations among authors in different countries for the measurement. The data was harvested from the *Science Citation Index* [46, 47]. Note that the rank-order was stable for two different years: $r = 0.98$ ($p < 0.001$).

**Table 5.** Triple Helix index for various countries and regions in the *Science Citation Index* [47].

| Countries | 2000 $\mu^*_{uig}$ in mbits | Rank | Countries | 2002 $\mu^*_{uig}$ in mbits |
|---|---|---|---|---|
| Japan | -92.1 | 1 | Japan | -82.4 |
| India | -78.1 | 2 | USA | -71.0 |
| USA | -74.4 | 3 | India | -67.7 |
| UK | -63.1 | 4 | UK | -54.0 |
| France | -52.1 | 5 | EU-15 | -45.3 |

**Table 5.** Cont.

| | | | | |
|---|---|---|---|---|
| EU-15 | -50.1 | 6 | France | -42.5 |
| Germany | -43.4 | 7 | Germany | -39.6 |
| S Korea | -40.1 | 8 | S Korea | -33.7 |
| Scandinavia | -31.6 | 9 | Netherlands | -32.8 |
| Italy | -29.4 | 10 | Scandinavia | -32.5 |
| Netherlands | -25.4 | 11 | Singapore | -28.6 |
| Russia | -24.2 | 12 | Italy | -27.6 |
| Singapore | -23.9 | 13 | Brazil | -26.8 |
| Brazil | -22.4 | 14 | Russia | -18.9 |
| Taiwan | -17.1 | 15 | Taiwan | -18.0 |
| PR China | -14.9 | 16 | PR China | -11.0 |

Table 5 suggests a very different pattern for the Triple Helix developments in various world regions. The Triple Helix overlay operates within the U.S.A. and Japan at a much higher level of self-organization than in Europe. Within the European Union, one can observe a scale with the U.K. at the leading end, but the smaller units (e.g., The Netherlands) at much lower levels. Russia and Brazil are even less integrated from this perspective, but India exhibits the Asian-Pacific pattern ($\mu^* < -0.70$). Interestingly, the Triple Helix overlay within People's Republic of China operates at a much lower level of self-organization than in other world regions.

**Figure 9.** Configurational information for South Korea and the Netherlands, respectively [47].

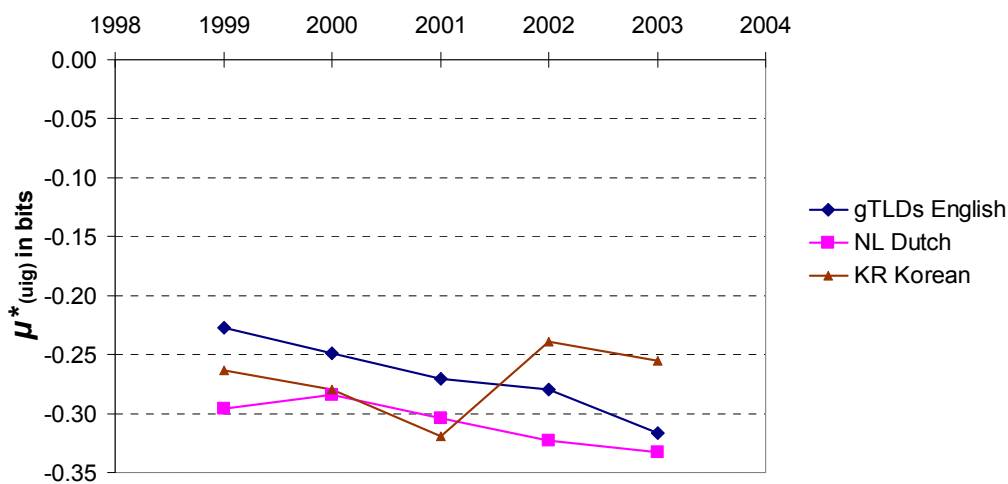

Figure 9 shows how a similar exercise measuring (co-)occurrences of the three terms ("university," "industry," and "government") at the Internet using the *AltaVista* Advanced Search Engine (on April 26, 2004) enabled us to indicate how the Korean domain was affected by the Internet crisis of 2000-2001, while the Dutch and generic Top Level Domains were not affected in terms of further strengthening these knowledge-based systems.

## 6. Configurational information and the dimensionality of the model

Shannon-type information is formal and does not inform us about external reality until the latter is specified as the system of reference for the application of the formal apparatus. The reduction of uncertainty in a potentially fourth dimension of the uncertainty, therefore, is not a property of reality, but a property of our models. The dimensionality of our models enables us to analyze increasing complexity.

While in Newtonian physics the data was given naturally and therefore the focus was on the measurement of this data, in Darwin's evolutionary theory, the observable data was considered as historical variation and selection as "natural." Retention mechanisms require stabilization of some selection environments. Stabilizations can again be at variance, and therefore be considered as second-order variation. The possibility of globalization adds a second-order selection term to this second-order variation [23].

Whereas stabilizations can be expected to further develop (or perish) along a trajectory, globalizing systems can be unfolded as a regime. The regime rests on the competing trajectories, and may thus reduce uncertainty. Whether a feed-back or a feed-forward is generated in the emerging dimension remains an empirical question, and as we have seen above will depend on the (near-)decomposibility of the mutual relations among subsystems [48].

**Table 6.** Organization of concepts in relation to degrees of freedom in the probability distribution [7].

|  | **first dimension** | **second dimension** | **third dimension** | **fourth dimension** |
|---|---|---|---|---|
| **Operation** | variation | selection | stabilization, retention | globalization and self-organization |
| **Nature** | entropy; disturbance | extension; network | localized trajectory | identity or regime |
| **Character of operation** | probabilistic; uncertain | deterministic; structural | reflexive; reconstructive | globally organized; resilient |
| **Appearance** | instantaneous and volatile | spatial; multi-variate | historically contingent | emerging hyper-cycle |
| **Unit of analysis** | change in terms of relations | latent positions | stabilities during history | virtual expectations |
| **Type of analysis** | descriptive registration | multi-variate analysis | time-series analysis | non-linear dynamics |

Each source of variance can be considered as a one-dimensional vector, but the subdynamics select upon each other's variation according to the systemic structure, that is, the redundancy, in their corresponding dimensions. Some selections can further be selected for stabilization along a trajectory, and stabilizations can be selected as second-order variation for globalization as a regime. In Table 6, I provide an overview of the relevant semantics in these four dimensions of probabilistic entropy [7].

## 7. Conclusions

Configurational information is generated when more than two sources of variance interact. Mutual information between two sources of variance is necessarily positive, but configurational information among three or more sources can be positive or negative. For example, when two parents have a child, and are able to structure the family situation, uncertainty is reduced for this child beyond his/her control. Obtaining an answer to a question asked of one parent improves the child's prediction about the answer the other parent can be expected to give. Thus, the situation is structured. Similarly, it makes a difference for government policies whether university-industry relations in a country are in good shape or whether they are in need of restructuring. However, the synergy in relations among three parties cannot be steered by any one of the participating parties. This synergy is the result of self-organization among the sources of variance.

The configurational outcome is more than the sum of its parts: it may contain less uncertainty. A new value emerges along an axis different from the three participating systems. The three participants relate and interact along analytically orthogonal axes, and a fourth axis emerges which may feed back (or forward) on the uncertainty which is thus generated. The resulting indicator can be used for measuring self-organization in the interactions among fluxes of communication in different dimensions. When three (or more) dimensions are relevant to the research design, configurational information can be measured, as in a Triple Helix of university-industry-government relations. The effects of interventions on changes in the values of parameters can be evaluated by using simulations.